\documentclass[twocolumn,letterpaper,aps,prc,superscriptaddress,nofootinbib,showpacs,floatfix]{revtex4-1}
\usepackage{graphicx}
\usepackage{comment}
\usepackage{setspace}
\usepackage[fleqn]{amsmath}
\usepackage[textwidth=18cm,textheight=24cm]{geometry}
\usepackage{color}
\usepackage{indentfirst}
\usepackage{xspace}
\usepackage{hyperref}
\usepackage{verbatim}
\usepackage{epstopdf}
\usepackage{bm}
\usepackage{csquotes}
\usepackage{multirow}

\usepackage{lineno}

\renewcommand{\Re}{\mathop{\mathrm{Re}}\nolimits}

\begin{document}
\title{Pion-Kaon femtoscopy as a probe of the space-time emission anisotropies due to interactions at the hadronic stage of matter evolution in relativistic heavy-ion collisions}
\author{P.~Chakraborty} 
\email{pritam.chakraborty@cern.ch}
\affiliation{Warsaw University of Technology, Faculty of Physics, ul. Koszykowa 75, 00-662, Warsaw, Poland}
\author{G.~Kornakov}
\email{georgy.kornakov@pw.edu.pl}
\affiliation{Warsaw University of Technology, Faculty of Physics, ul. Koszykowa 75, 00-662, Warsaw, Poland}
\author{A.~Kisiel}
\email{adam.kisiel@pw.edu.pl}
\affiliation{Warsaw University of Technology, Faculty of Physics, ul. Koszykowa 75, 00-662, Warsaw, Poland}
\author{Yu.~M.~Sinyukov}
\email{yu.sinyukov@gmail.com}
\affiliation{Warsaw University of Technology, Faculty of Physics, ul. Koszykowa 75, 00-662, Warsaw, Poland}
\affiliation{Bogolyubov Institute for Theoretical Physics, 14b Metrolohichna street, Kyiv 03143, Ukraine}
\author{V.~M.~Shapoval}
\email{shapoval@bitp.kyiv.ua}
\affiliation{Bogolyubov Institute for Theoretical Physics, 14b Metrolohichna street, Kyiv 03143, Ukraine}
\author{S.~Dash}
\email{sadhana@phy.iitb.ac.in}
\affiliation{Indian Institute of Technology Bombay, Powai, Mumbai - 400076, Maharashtra, India}


\date{\today}  


\begin{abstract}

Emission asymmetries between pions and kaons reflect the role of the hadronic phase in the cooling of a droplet of deconfined strongly-interacting matter.
This study compares results from two models at the same collision energy of $\sqrt{s_{\mathrm{NN}}}=5.02$~TeV to investigate how interactions in the hadronic phase affect particle emission. 
The first model, iHKM, provides a complete description of all stages of the evolution; from the initial scattering and thermalization to the final hadronic state, while the second model, LQTH (LHYQUID+THERMINATOR2), assumes a sudden conversion into hadrons, neglecting further interactions.
To increase the sensitivity to hadronic interactions, the analysis was performed as a function of the pair transverse velocity for pairs with nearly equal velocity vectors. 
The obtained predictions are compared with previously measured ALICE data at $\sqrt{s_{\mathrm{NN}}}=2.76$~TeV as a function of the cube root of the average particle multiplicity density at midrapidity, showing that both radii and emission asymmetries scale with particle multiplicity, regardless of the collision energy. 
The iHKM model reproduces the measured trends both qualitatively and quantitatively, whereas the LQTH model requires additional parameters—in particular, a time delay in the emission of kaons—to achieve quantitative agreement. 
The comparative analysis also indicates a possible non-monotonic behavior of the asymmetry as a function of transverse velocity, and a constant scaling of the ratio between the emission asymmetry and femtoscopic radii with particle multiplicity. 
These results highlight the importance of including interactions in the hadronic stage for a complete description of the emission function.
\end{abstract}

\maketitle


\section{Introduction} \label{intro}

The quark–gluon plasma (QGP), a state of strongly interacting matter where color charges are deconfined, is created at an intermediate stage of an ultra-relativistic A--A collision.
Since the QGP behaves as a nearly perfect fluid with the lowest known shear viscosity ($\eta/s\approx 0.1$)~\cite{qgp1,qgp2}, various soft-physics hadronic observables measured in heavy-ion collision experiments,
such as transverse momentum ($p_{\mathrm{T}}$) spectra, anisotropic flow coefficients $v_2$, femtoscopy radii $R_i$ and related observables, can be accurately described using models based on relativistic hydrodynamic calculations. 
Such calculations require assuming an equation of state for the QGP that, at zero baryon chemical potential, is consistent with lattice QCD predictions of a crossover transition between quark-gluon and hadronic phases. 
The collective flow should start developing in the model as early as possible, already at the prethermal stage of the matter evolution,
right after the initial nuclear overlap (at proper times $\tau \sim 0.1$~fm/$c$). 
Additionally, resonance decays should also be taken into account. As for the role of hadronic interactions at the final stage of a collision in the formation of observables, there are two widely used approaches. 
The ``sudden freeze-out'' approach used in models with statistical hadronization suggests an almost absent hadronic stage, with a single freeze-out of particle spectra and the system's chemical composition at the hadronization hypersurface (plus accounting for resonance decay contribution). 
The ``continuous freeze-out'' approach, typical for hybrid models that include the hadronic cascade stage, assumes a significant influence of this final stage on the resulting observables.

Correlation femtoscopy~\cite{goldhaber,kopylov1,kopylov2,cocconi}, an experimental technique based on measurements of two-particle momentum correlations, has been used for decades to estimate the spatiotemporal dimensions of extremely small, hot and dense systems formed in relativistic heavy-ion collisions. 
The method presents the spatial and temporal extents of the system in terms of \textit{femtoscopy radii} extracted from the Gaussian fits to the corresponding correlation functions and associated with the \textit{homogeneity lengths}, i.e. the sizes of the regions from which particles with a given mean velocity are mostly emitted~\cite{gyulassy,pratt-hbt,makhsin87,makhsin88,sin89,hama-hbt,sin94,akksin95}.

The results of identical pion femtoscopy measurements in heavy-ion collisions at the  RHIC and LHC energies show that the femtoscopic radii decrease with increasing average pair transverse momentum, $\mathbf{k}_{\mathrm{T}}=[\mathbf{p}_{\mathrm{1,T}}+\mathbf{p}_{\mathrm{2,T}}]/2$, and pair transverse mass, $m_{\mathrm{T}}=\sqrt{k^2_{\mathrm{T}} + m^2}$, indicating that the source dimensions, or more accurately, the homogeneity lengths are significantly influenced by the collective flow.  
Based on simple hydrodynamic considerations, one could also expect
the radii for different particle pairs to follow universal $m_{\mathrm{T}}$-scaling behavior. 
However, the comparison of
identical pion and kaon femtoscopy radii dependencies on $m_\mathrm{T}$
in the LHC Pb$-$Pb collisions reveals a clear breaking of the $m_{\mathrm{T}}$-scaling~\cite{kaon_2760}. 
This breaking was predicted based on the hydrokinetic model (HKM) calculations~\cite{hkm-kaon} and
is well reproduced by other hybrid models which account for the transverse collective motion of the matter at the hydrodynamic stage of the system's evolution (at least in (2+1)D boost-invariant regime) and for the hadronic rescatterings at the final ``afterburner'' phase following the system's hadronization. 
Thus, the final hadronic phase appears to have a strong influence on the observed femtoscopy scales.

Over the past few years, the femtoscopy of non-identical particle pairs has attracted considerable attention in both heavy-ion and hadronic collision studies. 
It allows one to probe the source geometry and particle-emission dynamics for particles of different species and to investigate the final-state interactions (FSI) between them.
Since pairs of non-identical particles are considered, it avoids the need to disentangle the problem of separating quantum-statistics correlations from those caused by FSI (strong and Coulomb interactions). 

One of the major perks of non-identical particle femtoscopy is that it provides the means to measure the spatial difference between mean emission points of two considered particle species, 
referred to as \enquote{pair emission asymmetry}. 
The 
concept of this asymmetry
and the method to estimate it from the two-particle momentum correlation functions were first suggested in~\cite{led1,led2} and then discussed in detail in~\cite{adam_nonid}, where the systems produced in $\sqrt{s_{\mathrm{NN}}}=200$~GeV nucleus-nucleus collisions are considered. 
The calculations show that the interplay between the radial flow and thermal velocity in a hydrodynamically expanding system produces the asymmetry in the average emission points of the particles, so that pions are emitted later in time and closer to the system's center than kaons or protons. 
Moreover, the analysis of pion-kaon and pion-proton correlations shows that this asymmetry is enhanced by the contribution of resonance decay products. However, the maximal emission time $\tau$ estimation for pions and kaons within HKM, carried out in~\cite{mtscales1} and later in~\cite{mtscales2,mtscales3}, in contrast, gave $\tau_K>\tau_\pi$.
This result was then confirmed in the ALICE Collaboration's experimental analysis~\cite{kaon_2760} and attributed to the complex afterburner dynamics including rescattering and recombination of $K^*(892)$ resonance decay products, $K^*(892) \leftrightarrow K\pi$. 
The $K^*$ decays serve as
a rich source of secondary kaons, and a significant fraction of observed $K^*$'s (about 60\%) is produced via recombination of kaon-pion pairs
at late times of the system's evolution~\cite{kstar}, so that hadronic interactions at the afterburner stage can result in later kaon emission as compared to pions. 
Pions are more abundantly produced from other sources, so pions coming from $K^*$ decays, while delayed in exactly the same way as kaons, are a relatively small fraction, and have only a weak influence on the overall average pion emission time.

The results of pion-kaon femtoscopic analysis in Pb$-$Pb collisions at $\sqrt{s_{\mathrm{NN}}}=2.76$~TeV by the ALICE experiment also show that the measured pair-emission asymmetry is significantly influenced by the combined effect of the radial flow and hadronic rescattering phase~\cite{pionkaon_2760_alice}. 
Since so far these experimental results were compared only with the predictions from the (3+1)D hydrodynamics+THERMINATOR2 model, which does not include the rescattering phase explicitly, 
it is therefore instructive to compare also the measured data with the predictions from a realistic collision model that incorporates hadron rescattering effects, like the integrated Hydro-Kinetic Model (iHKM)~\cite{ihkm2,ihkm}, an improved version of HKM. 
Moreover, it is also essential to investigate the variation of pair-emission asymmetry as a function of collective velocity for a deeper understanding of pair-emission dynamics. 
This can be achieved by analyzing
pair-emission asymmetry for different ranges of mean pair velocity $\beta_\mathrm{T}$.

In this work, 
we present results from 
a comparative analysis of the correlation functions of charged pion-kaon pairs calculated in the two models, iHKM and LHYQUID+THERMINATOR2 (LQTH), 
for Pb$-$Pb collisions at the LHC energy $\sqrt{s_{\mathrm{NN}}}=5.02$~TeV. 
The predicted source parameters are also compared with the pion-kaon femtoscopy results from the ALICE experiment for Pb$-$Pb collisions at a lower collision energy of  $\sqrt{s_{\mathrm{NN}}}=2.76$~TeV. 
It is worth noting that in this paper, we analyze, for the first time, the correlation functions in different pair transverse velocity ($\beta_{\mathrm{T}}$) ranges, using two different models in order to directly observe the effects of 
collective flow and particle interactions during the hadronic phase on the extracted femtoscopic source parameters (radii and asymmetries). 

The paper is organized as follows. In Section~\ref{sec:model}, a brief description of the collision models used in this analysis is given. The formalism of the correlation function construction and its fitting procedure are explained in Sections~\ref{sec:CF_th} and \ref{sec:fit}, respectively. Section~\ref{sec:primfrac} contains an extensive discussion on purity, the fraction of femtoscopically correlated pairs, which is an important parameter used in the fitting of the correlation function. The results of the analysis are presented and discussed in Section~\ref{sec:results}. Finally, Section~\ref{sec:conclusion} presents the conclusions of our analysis and its implications for experimental studies.

\section{Models' description} \label{sec:model}

A brief description of the two models used in this paper to simulate ultrarelativistic heavy-ion collisions, iHKM and LQTH, is provided below. 
Both models treat the collision process as consisting of several successive stages, and thus provide a realistic description of the evolution of the created system.

The iHKM model includes the following five stages: initial state formation, prethermal relaxation dynamics, hydrodynamic expansion, particlization, and hadronic cascade. 
In the first stage, corresponding to the initial proper time $\tau_0 \approx 0.1$~fm$/c$ after the collision of two heavy ions, 
the system is assumed to be in a non-equilibrated partonic state with an initial energy density profile, simulated using Monte Carlo (MC) Glauber calculations within the GLISSANDO code~\cite{gliss}, and an anisotropic momentum distribution inspired by the Color Glass Condensate model~\cite{cgc1,cgc2}. 
During the prethermal stage, the system gradually evolves from the initial non-equilibrated state to a nearly locally equilibrated one, 
consistent with its subsequent hydrodynamic treatment as a continuous medium. 
Thermalization is assumed to occur at $\tau_{th}\approx 1$~fm$/c$. The following stage involves relativistic viscous (2+1)D boost-invariant hydrodynamics expansion ($\eta/s=0.08$), described within the Israel-Stewart formalism~\cite{is3}. 
The hydrodynamic equations are solved numerically using the HLLE code~\cite{hlle}.
A lattice-QCD-inspired equation of state is employed to characterize the properties of quark-gluon matter at this stage~\cite{laine,hotqcd}.
As the system expands and cools, it reaches the \enquote{particlization temperature} $T_p \approx 160$~MeV, where the hydrodynamic description ceases to be valid. 
The system is then converted into a set of hadrons according to the Cooper-Frye prescription~\cite{cooper-frye}, with viscosity corrections~\cite{grad,cornelius3}, implemented via the Cornelius routine~\cite{cornelius3,cornelius1}. 
In the final stage, the UrQMD hadron cascade code~\cite{urqmd1,urqmd2} is used to simulate multiple elastic and inelastic scatterings within the hadronic phase, as well as resonance propagation and their decays.

In the LQTH model~\cite{lqth1,lqth2}, the (3+1)-dimensional viscous hydrodynamics (LHYQUID) describes the collective dynamics of the system created in a heavy-ion collision by solving the second-order Israel-Stewart equations, assuming viscosity coefficients $\zeta/s=0.04$ and $\eta/s=0.08$, with all chemical potentials set to zero~\cite{lqth1}. 
The hydrodynamic expansion begins at $\tau=0.6$~fm$/c$ and continues until  the single freeze-out, i.e. the simultaneous chemical and kinetic freeze-out of the system, which occurs at the temperature $T_{\rm f}=140$~MeV. 
At this temperature, the system hadronizes, with the process implemented using the THERMINATOR2 package~\cite{lqth2}, 
based on the Cooper-Frye formalism for statistical hadronization at the freeze-out hypersurface. 
THERMINATOR2 is a Monte Carlo event generator developed using the CERN ROOT C++ framework. 
This package does not include hadronic rescatterings but simulates the propagation and decay of resonances. 


\section{Non-identical-particle femtoscopy methodology} \label{sec:CF_th}
The two-particle momentum correlation function (CF) for non-identical particles is defined as the ratio of the probability $P_{\mathrm 1,2}(p_1,p_2)$ of detecting two particles  
with momenta $p_{\rm 1}$ and $p_{\rm 2}$  
to the product of corresponding single-particle probabilities  
$P_{\mathrm 1}(p_i)$ and $P_{\mathrm 2}(p_i)$~\cite{adam_nonid}:
\begin{eqnarray}
	C(p_{\rm 1}, p_{\rm 2}) = \frac{P_{1,2}(p_{\rm 1}, p_{\rm 2})}{P_1(p_{\rm 1})P_2(p_{\rm 2})}. 
    \label{eq1}  
\end{eqnarray}

In experiment, the two-particle correlation function is constructed by estimating the ratio of the correlated pairs distribution, referred to as the \textit{Signal} ($T$), to the distribution of uncorrelated pairs, referred to as the \textit{Background}~($M$): 
\begin{eqnarray}
 C(\mathbf{k}^*)=T(\mathbf{k}^*)/M({\mathbf{k}^*}). 
 \label{eq:expt_C}
\end{eqnarray}
The \textit{Signal} is constructed by selecting the particle pairs, where both particles come from the same event, whereas the \textit{Background} is constructed by pairing the particles selected from different events.

The Pair Rest Frame (PRF), where total momentum of the pair vanishes, is of special importance in correlation analysis. In this work, values in PRF are denoted with an asterisk ($^{*})$.
The correlation is constructed as a function of half of the pair relative momentum $k^{*}=(p^{*}_1-p^{*}_2)/2$. For a pair of identical particles $k^{*}=q_{inv}/2$ \cite{Therminator_identical_2760MeV}, while for pairs containing different particles, such as pion-kaon pairs considered in this work, $k^{*}$ corresponds to the momentum of the first particle in the pair in PRF. In the current study we consistently use the convention where pions are taken as ``the first'' in the pair.
The dependence on the total pair momentum $P$ is investigated by considering the correlation functions $C_P(k^{*})$ at different fixed $P$ intervals, and such study is the focus of this work.

For the quantitative analysis using the theoretical models, within smoothness, on-shell and equal-time approximations 
(see the details and the derivation from Eq.~(\ref{eq1}) in, e.g., \cite{source}), 
the correlation function can be connected to 
the pair relative wave function $\Psi(\mathbf{r}^*,\mathbf{k}^*)$ and the \textit{source function} $S_\mathbf{P}(\mathbf{r}^*)$ --- in the so-called Koonin-Pratt approach~\cite{koonin0,koonin1,koonin2,koonin3}:
\begin{eqnarray}	 
	C_\mathbf{P}(\mathbf{k}^*) = \int |\Psi(\mathbf{r}^*,\mathbf{k}^*)|^2S_{\mathbf{P}}(\mathbf{r}^*) d^3\mathbf{r}^*, 
    \label{eq:koonin}
\end{eqnarray}
where $\mathbf{r}^*$ is the relative space separation of the pair in PRF and $S_\mathbf{P}(\mathbf{r}^*)$ is the time-integrated separation
distribution at given total pair momentum $\mathbf{P}$.

In standard approach, to fit the experimental correlation function one firstly assumes a certain ``theoretical'' $S_\mathbf{P}(\mathbf{r})$, usually, a three-dimensional (3D) spheroid with Gaussian density profile, having different widths in the three directions. 
This simple source function is used to calculate (based on Eq.~(\ref{eq:koonin})) the ``theoretical'' correlation function, which is then used to fit the experimental CF.
Using the Bertsch-Pratt coordinate system
in the Longitudinally Co-Moving System of reference (LCMS), one commonly considers three spatial directions: \enquote{out} --- along the pair total transverse momentum, \enquote{long} --- coinciding with the beam direction, and \enquote{side}, which is perpendicular to the other two.  
In LCMS, the total momentum of pairs in the \textit{long} direction is zero by definition. 
Also, in a collider setup for heavy-ion collisions,
the asymmetry in the pair emission is expected to occur only in the \textit{out} direction~\cite{adam_nonid}.
The theoretical 
source function, $S_{\mathbf{P}}({{\mathbf{r}}})$ 
can then be expressed as follows:
\begin{equation}
	\label{eq:sourcefcn}
	S_{\mathbf{P}}({{\mathbf{r}}})=\exp{\left[-\frac{(r_{\rm out}-\mu_{\rm out,{\mathbf{P}}})^2}{2R^2_{\rm out,{\mathbf{P}}}}-\frac{r^2_{\rm side}}{2R^2_{\rm side,{\mathbf{P}}}}
	-\frac{r^2_{\rm long}}{2R^2_{\rm long,{\mathbf{P}}}}\right]},
\end{equation}
where, $R_{\rm out,{\mathbf{P}}}$, $R_{\rm side,{\mathbf{P}}}$ and $R_{\rm long,{\mathbf{P}}}$ are the source sizes in \textit{out}, \textit{side} and \textit{long} directions, respectively, and $\mu_{\rm out,{\mathbf{P}}}$ are the pair emission asymmetries along the \textit{out} direction at given total momentum $\textbf{P}$.

The function $\Psi$ in Eq.~(\ref{eq:koonin}), referred to as the pair relative wave function, corresponds to the interaction between  particles in a pair at given relative momentum and given pair separation. 
If both particles in the pair are identical it must be properly (anti-)symmetrized \cite{Therminator_identical_5020MeV}, 
while for pairs of non-identical particles, such as the pion-kaon pairs used in this work, only the FSI effect needs to be accounted for. Although for the charged pion-kaon pairs, the FSI includes both Coulomb and strong interaction, the latter is assessed to be relatively small.
In this theoretical study it is sufficient to consider only the Coulomb interaction, provided that the same approach is self-consistently used both in the calculation of "model" correlation functions and in their subsequent fitting procedure. While it is technically possible to include full Coulomb+Strong FSI in the theoretical calculations, it is significantly more computationally expensive and does not qualitatively change the result, so in this work only the Coulomb FSI is used.
The wave function can the be expressed as~\cite{Lednicky:2005tb}: 
\begin{eqnarray}
	\Psi(\mathbf{r}^*, \mathbf{k}^*)=\sqrt{A_C(\eta)}\left[e^{-i\mathbf{k}^*\mathbf{r}^*}F(-i\eta,1,i\zeta)\right],  \label{eq:wave}
\end{eqnarray}
where, $A_C$ is the Gamow penetration factor, $\eta=1/(k^*a_c)$, $a_c$ is the Bohr radius of the pair, equal to $\pm 248.58$~fm for like- and unlike-charged pion-kaon pairs respectively, $F$ is the confluent hypergeometric function, $\zeta=k^*r^*(1+\cos\theta^*)$, $\theta^*$ is the angle between $\mathbf{k}^*$ and $\mathbf{r}^*$. 

In the theoretical models available for heavy-ion collisions, one typically does not implement FSI directly on microscopic level. Instead, the FSI effect is artificially 
imposed on particle pairs after the hadronization using the \enquote{afterburner} procedure, where the $\mathbf{k}^*$ of each pair, selected from the same events is added to the $T(\mathbf{k}^*)$ distribution with the weight of $|\Psi(\mathbf{r}^*,\mathbf{k}^*)|^2$, calculated using Eq.~(\ref{eq:wave})~\cite{adam_nonid}. 

In this work, correlation functions are stored and analyzed in terms of spherical harmonics (SH). The procedure of calculating the SH components of the correlation functions is thoroughly discussed in \cite{adamsh}. It was shown in \cite{adam_nonid} that using only two SH components of the correlation function, $C^{\mathrm0}_{\mathrm0}$ and $\Re C^{\mathrm1}_{\mathrm1}$, is enough to probe the size of the \enquote{region of homogeneity} as well as the pair-emission asymmetry along the \textit{out} direction. 

In this study, we analyze the iHKM and LQTH simulation results for $C^{\mathrm0}_{\mathrm0}$ and $\Re C^{\mathrm1}_{\mathrm1}$ components of the correlation functions of charged pion-kaon pairs in Pb$-$Pb collisions at $\sqrt{s_{\mathrm {NN}}}=5.02$~TeV. 
Five centrality classes (0$-$5\%, 10$-$20\%, 20$-$30\%, 30$-$40\% and 40$-$50\%) are addressed with the pseudo-rapidity cut $|\eta|<0.8$ (to match the acceptance constraints of the ALICE detector). 
The following two variants of simulated data selection are considered: 
\begin{itemize}
    \item pairs of particles with $0.19<p_{\mathrm T}<1.50$~GeV$/c$ are selected for a direct comparison with the ALICE Collaboration experimental results for Pb$-$Pb collisions at $\sqrt{s_{\mathrm {NN}}}=2.76$~TeV.
    \item pairs made of pions and kaons with $0.1<p_{\mathrm T}< 2.5$~GeV$/c$ and $0.2<p_{\mathrm T}<2.5$~GeV$/c$, respectively, are divided into the following $\beta_{\mathrm T}$ intervals: 0.70$-$0.75, 0.75$-$0.80, 0.80$-$0.85, 0.85$-$0.90 and 0.90$-$0.95.
\end{itemize}
The $C^{\mathrm0}_{\mathrm0}$ and $\Re C^{\mathrm1}_{\mathrm1}$  components of example $\pi^+K^+$ and $\pi^+K^-$ correlation functions calculated in the iHKM model are shown in Fig.~\ref{fig:femto_corr}. The $C^{\mathrm0}_{\mathrm0}$ components deflect from 1 in the low-$k^*$ region due to the attractive and repulsive Coulomb interaction. The deviation of $\Re C^{\mathrm1}_{\mathrm1}$ from 0 indicates the presence of asymmetry in the pair emission.


\section{Fitting of the correlation function} \label{sec:fit}
In order to compare the model calculations performed in this work with experimental data, both should be treated in the same manner, as closely as possible. Therefore, the correlation functions calculated according to the procedure described in Sec.~\ref{sec:CF_th} are treated with the same fitting procedure, which would be applied to the ones from experiment, with the exception that, in this work only Coulomb interaction is included in the FSI calculation, while for real data full Coulomb+Strong FSI would be used.

The fitting procedure relies on the numerical integration of Eq.~\eqref{eq:koonin}, and has been explained extensively in~\cite{adam_nonid, Therminator_Nonidentical_2760MeV}. The chosen source parameterization, given by Eq.~\eqref{eq:sourcefcn}, is justified in LCMS, rather than in PRF. 
This reference frame is routinely used in analysis of three-dimensional femtoscopy of charged pions and kaons, as it provides the most straightforward interpretation of femtoscopic radii in relation to the reference frame of the fluid element emitting particles.
In its general form, Eq.~\eqref{eq:sourcefcn}
has three radii and the pair-emission asymmetry as the free parameters in the fitting procedure. However, it was observed in the previous studies that 
the $C_{0}^{0}$ and the $\Re C_{1}^{1}$ components of the correlation function are only sensitive to the overall, directionally-averaged source size and the emission asymmetry, and are not sufficiently sensitive to the details of the 3D structure of the source.
Hence, 
to reduce the number of fitted parameters 
the following assumptions, based on the results from identical pion femtoscopy in heavy-ion collisions from the STAR and the ALICE experiments were used: $R_{\rm side}=R_{\rm out}$ and $R_{\rm long}=1.3R_{\rm out}$. As a result, only two independent parameters, $R_{\rm out}$ and $\mu_{\rm out}$, remain in the fitting procedure.

       \begin{figure}[!h]
				 \centering
         {\includegraphics[width=0.5\textwidth]{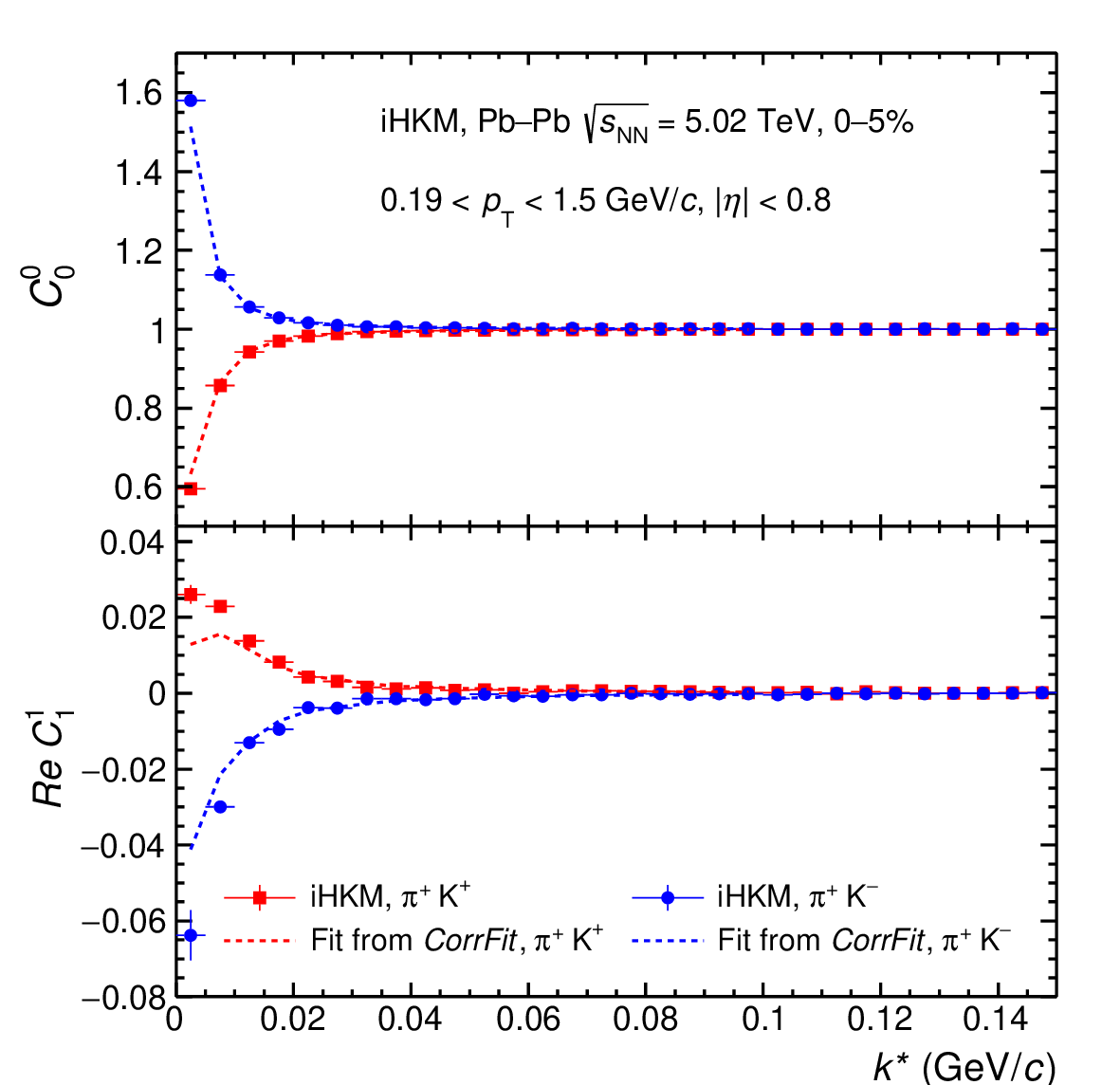}}
				 \caption{(Color online) Spherical Harmonic components ($C^0_0$ and $\Re C^1_1$) of the femtoscopic correlation function of charged pion-kaon pairs calculated in iHKM for 0-5\% central Pb–Pb collisions at $\sqrt{s_{\rm NN}}=5.02$~TeV with the respective fit lines estimated in the $CorrFit$ package.}
				  \label{fig:femto_corr}
		\end{figure}
        
It was observed in \cite{adam_nonid} that the source-size parameters affect all the SH components simultaneously, which compels one to fit the components relevant in this work, i.e., $C^{\mathrm0}_{\mathrm0}$ and $\Re C^{\mathrm1}_{\mathrm1}$, together. 
No analytical form of the fitting function exists for Coulomb FSI in SH. 
The correlation functions are fitted with a numerical procedure using the \textit{CorrFit} software package.  
The values of the two parameters, $R_{out}$ and $\mu_{out}$, are selected. A sample of pairs is selected, with kinematics constraints as close as possible to the ones used to obtain the correlation function. In real data, true momenta of the sample of pairs is simply taken. In this work a sample of pairs is generated from the model.
For each pair the emission points are randomly generated according to Eq.~\eqref{eq:sourcefcn} with the given value of
$R_{\rm out}$ and $\mu_{\rm out}$.
With the momenta and emission points of both particles determined, the weight $w$, equal to the modulus squared of the pair relative wave function can be calculated according to Eq.~\eqref{eq:wave}. The signal histogram $T(k^{*})$ is then filled with $w$ at the corresponding $k^{*}$, while the background histogram $M(k^{*})$ is filled with 1.0. After the procedure has been repeated for all the pairs in the sample, the $T$ and $M$ histograms are used to calculate the ``fit'' correlation function according to Eq.~\eqref{eq:expt_C}. The whole procedure is repeated for a number of $(R_\mathrm{out}, \mu_\mathrm{out})$ pairs forming a grid in a selected range of values for both parameters. At each grid point, the ``fit'' function (calculated as described above) is compared to the ``experimental'' correlation function (calculated from the model) via the $\chi^{2}$ test. The result is a two-dimensional $\chi^{2}$ map, for which a minimum can be found. The location of this minimum is the result of the fit. 

Fig.~\ref{fig:femto_corr} shows the outcome of the fitting procedure. The points represent the ``experimental'' correlation functions, which are being fitted (in this case, the CFs calculated for $\pi^+K^+$ and $\pi^+K^-$ pairs from the iHKM model at the selected centrality). 
The dashed lines show the ``fitted'' correlation function, calculated with the procedure described above, for the ``best-fit'' $R_\mathrm{out}$ and $\mu_\mathrm{out}$ values. 
The fitting procedure has one additional complication, related to the so-called
\enquote{fraction of femtoscopically correlated pairs}, which is discussed in detail in the next Section.


\section{Fraction of femtoscopically correlated pairs} \label{sec:primfrac}

\begin{figure}[!h]
    \centering    {\includegraphics[width=0.5\textwidth]{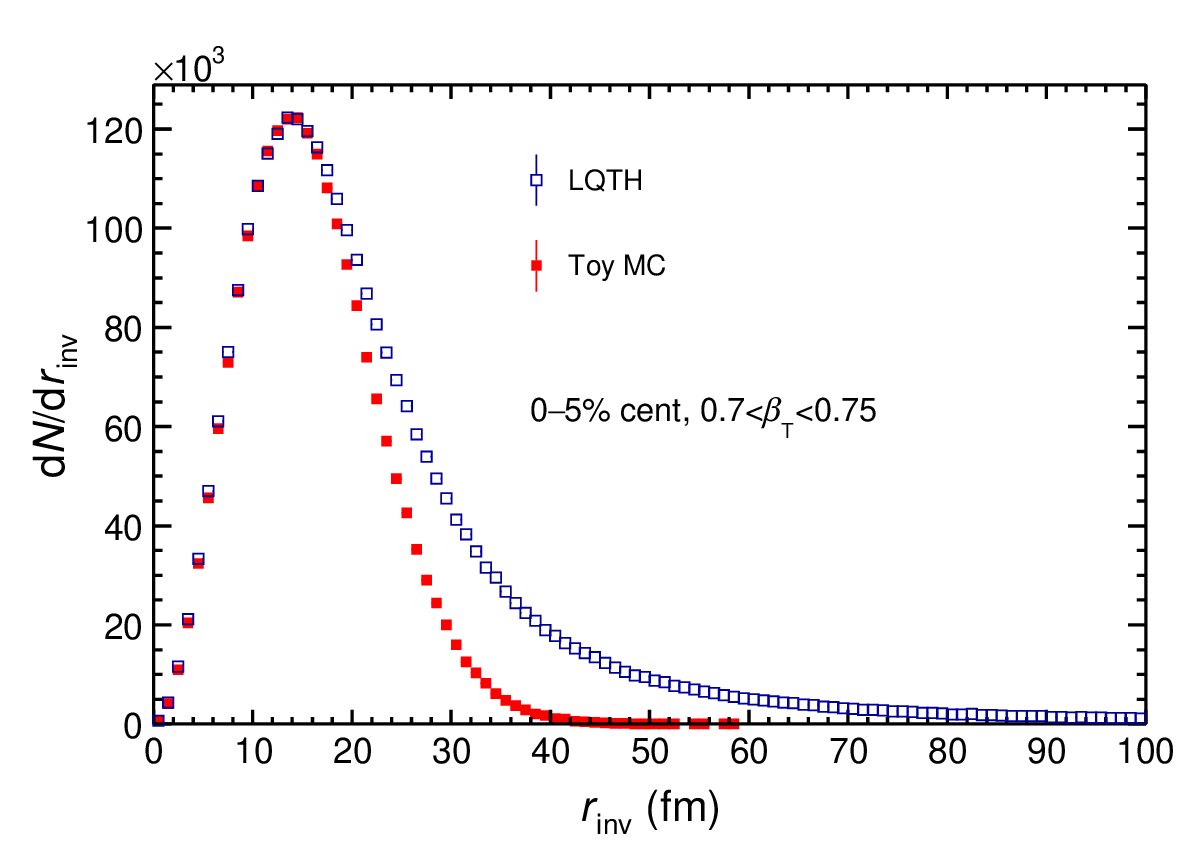}}
    \caption{The $r_{\rm inv}$ distribution with long-range tails from the resonance decays (blue) and the corresponding Gaussian fit function (red).} \label{fig:femto_rinv}
\end{figure}

The source function, as discussed in Section~\ref{sec:CF_th}, is assumed to be a Gaussian one, however, the model calculations in \cite{adam_nonid} show the presence of non-Gaussian long-range tail in the 1D distribution $dN/dr_{\rm inv}$ corresponding to the original 3D source function. In Fig.~\ref{fig:femto_rinv}, the $r_{\rm inv}$ distribution (blue points) for $\pi^+K^-$ pairs selected within the range $0.70<\beta_{\rm T}<0.75$ in 0$-$5\% central events, calculated directly from LQTH model, is shown together with the Gaussian fit to it (red points). The non-Gaussian tail is clearly visible upon comparison of the actual distribution with the purely Gaussian one.
The particles produced from strongly decaying long-lived resonances contribute mostly to this long-range tail. Since the FSI between the particles is maximum in pairs at very small separation, the pairs from the non-Gaussian tail do not contribute to the femtoscopic correlation. Hence, one can estimate the fraction of femtoscopically correlated pairs, $f_{G}$, as the ratio of the number of particle pairs coming only from the Gaussian core of the source function to the number of all pairs. 
The effect of the $f_{G}$ value on the correlation function is mathematically equivalent to the ``purity'' factor commonly applied in femtoscopic analyses. Effectively, the pairs within the Gaussian core contribute to the correlation with their full weight, while pairs outside of the core do not contribute to the correlation effect at all. In order to correct for this phenomenon, the theoretical ``fit'' correlation function $C_\mathrm{calc}$ should be corrected as follows:
\begin{equation} 
    C_\mathrm{corr} = 1 + (1 - C_\mathrm{calc})/f_{G},
    \label{eq:purcor}
\end{equation}
and the corrected function $C_\mathrm{corr}$ should be used in the $\chi^{2}$ tests described in the previous section. Eq.~\eqref{eq:purcor} shows that the correction directly affects the magnitude of the correlation, which is most sensitive to the size of the system. Therefore, the correct estimation of the value of this parameter is crucial for the proper fitting procedure. 

The simplest way to calculate $f_{G}$ is to fit the $r_{\rm inv}$ distribution with a Gaussian function 
\begin{eqnarray}
F_G(r_{\mathrm{inv}}) \sim r^2_{\mathrm{inv}}\exp\left(-\frac{r^2_{\mathrm{inv}}}{2R^2_{\mathrm{avg}}}\right), \label{eq:gauss_1d}
\end{eqnarray}
where $R_{\mathrm{avg}} = \sqrt{(R^2_{\mathrm{out}}+ R^2_{\mathrm{side}} + R^2_{\mathrm{long}})/3}$. However, this procedure is exact only when the radii in all three directions are equal --- a condition which is not satisfied in this study, as discussed in Section~\ref{sec:fit}. Hence, to calculate $f_{G}$ more accurately, a Toy Monte Carlo model with a three-dimensional source function, perfectly Gaussian in LCMS, is considered.  
The relative pair separation projections along the three axes, boosted to the PRF, can be randomly sampled using the Gaussian distributions. Generating the $r^*_{\mathrm{side}}$ and $r^*_{\mathrm{long}}$ components is pretty straightforward, since $r^*_{\mathrm{side}} = r_{\mathrm{side}}$, $r^*_{\mathrm{long}}=r_{\mathrm{long}}$, and the respective distributions 
do not have any shifts, due to the symmetry in heavy-ion collisions: 
\begin{eqnarray} \label{eq:toy_side}
    r^*_{\mathrm{side}} = r_{\mathrm{side}} \sim {\it G} (0,\sigma_{\mathrm{side}}), \nonumber\\
    r^*_{\mathrm{long}} = r_{\mathrm{long}} \sim {\it G} (0,\sigma_{\mathrm{long}}),
\end{eqnarray}
where $G(r_{0i},\sigma_i)$ corresponds to random-number generation according to  the Gaussian distribution with the mean values $r_{0i}$ and widths $\sigma_i$. 
In case of $r_{\mathrm{out}}$, the distribution has the mean value shifted by $-\mu_{\mathrm{out}}$ and the width $\sigma_{\mathrm{out}}$, both scaled by the Lorentz-factor $\gamma_\mathrm{T}$, due to transformation from LCMS to PRF.
Thus, the distribution of $r^*_{\mathrm{out}}$ can be generated as
\begin{eqnarray}
    r^*_{\mathrm{out}}\sim {\it G} (-\gamma_{\mathrm T}\mu_{\mathrm{out}},\gamma_{\mathrm T}\sigma_{\mathrm{out}}),\label{eq:toy_out}
\end{eqnarray}
where $\gamma_{\rm T} = \sqrt{1-\beta_{\rm T}^2}$, $\beta_{\rm T}$ is the pair transverse velocity. With the randomly generated values of $r^*_{\rm{out}}$, $r^*_{\rm{side}}$ and $r^*_{\rm{long}}$, the $r^*\equiv r_{\mathrm{inv}}$ can be calculated as follows:
\begin{eqnarray} \label{eq:r_PRF}
        r_{\mathrm{inv,toy}} = \sqrt{{r^*}^2_{\mathrm{out}}+{r^*}^2_{\mathrm{side}}+{r^*}^2_{\mathrm{long}}}.
\end{eqnarray}
The $\sigma$'s used in generating the pair-separation distributions 
must follow the same assumptions, as the ones used in the fitting procedure, i.e. $\sigma_{\mathrm{side}}=\sigma_{\mathrm{out}}$ and $\sigma_{\mathrm{long}}=1.3\sigma_{\mathrm{out}}$. While this approximation reduces the number of free parameters used in the calculation of $r_{\mathrm{inv}}$, unfortunately, no such approximation can be applied for $\mu_{\mathrm{out}}$. To resolve this issue, all the calculations related to the estimation of $f_G$ and the fitting of the correlation functions are performed in two steps. 

In the first step, the $r^*_{\mathrm{out}}$ and $r_{\mathrm{inv,toy}}$ distributions are calculated by assuming $\mu_{\mathrm{out}}=0$. Then the value of $\sigma_{\mathrm{out}}$ for which the $r_{\mathrm{inv, toy}}$ distribution matches with the $r_{\mathrm{inv}}$ from the model calculations in the lower pair separation region is estimated. The distributions of $r_{\mathrm{inv, toy}}$ (red points in Fig.~\ref{fig:femto_rinv}) and $r_{\mathrm{inv}}$ are integrated over the full available range and their ratio is estimated as the $f_G$. This process is performed for LQTH and iHKM separately in each centrality class and $\beta_{\mathrm T}$ interval to control the systematic variation of $f_G$. Another source of systematic variation is the primary fraction used as the input parameter in \textit{CorrFit} to fit the one-dimensional correlation function which produces the same $\sigma_{\mathrm{out}}$ as the toy MC. The average $f_G$ is estimated along with its uncertainty from these four pair charge variations in each centrality class and $\beta_{\mathrm T}$ interval, which are then used to fit the $C_{\mathrm 0}^{\mathrm 0}$ and $\Re C_{\mathrm 1}^{\mathrm 1}$ components from LQTH and iHKM to estimate the values of $R_{\mathrm{out}}$ and $\mu_{\mathrm{out}}$. Next, the average values of $\mu_{\mathrm{out}}/R_{\mathrm{out}}$ in each $\beta_{\mathrm T}$ are estimated and the $\mu_{\mathrm{out}}/R_{\mathrm{out}}$ vs. $\beta_{\mathrm T}$ distribution is fitted with a polynomial of order 3, which provides the approximated ratio of $\mu_{\mathrm{out}}$ to $R_{\mathrm{out}}$ in a given $\beta_{\mathrm T}$ interval. Although this approximation is very much model dependent, it provides a preliminary estimation of the dependence of $\mu_{\mathrm{out}}$ on $R_{\mathrm{out}}$.

In the next step, the whole process of estimating the $r_{\mathrm{inv,toy}}$ is repeated by using $\mu_{\mathrm{out}}$ and a scaled value of $\sigma_{\mathrm{out}}$ in each $\beta_{\mathrm T}$ interval, as calculated at the final step of the process mentioned in the last paragraph. It was observed that the fluctuations in the $f_G$ reduced significantly after incorporating non-zero values of the $\mu_{\mathrm{out}}$ in $r^*_{\mathrm{out}}$ estimation. The final values of $f_G$ in each $\beta_{\mathrm{T}}$ interval and centrality class are given in Tables~\ref{tab:fG_bT_1} and \ref{tab:fG_bT_2}.

        \begin{table}[!h]
        \begin{tabular}{|c|ccc|}
        \hline
        \multirow{2}{*}{$\beta_{\rm T}$} & \multicolumn{3}{c|}{Pair-purity for different centralities}                             \\ \cline{2-4} 
                                         & \multicolumn{1}{c|}{0-5\%}         & \multicolumn{1}{c|}{10-20\%}       & 20-30\%       \\ \hline
        0.70--0.75                         & \multicolumn{1}{c|}{0.73$\pm$0.05} & \multicolumn{1}{c|}{0.65$\pm$0.06} & 0.69$\pm$0.05 \\ \hline
        0.75--0.80                         & \multicolumn{1}{c|}{0.67$\pm$0.06} & \multicolumn{1}{c|}{0.66$\pm$0.04} & 0.66$\pm$0.04 \\ \hline
        0.80--0.85                         & \multicolumn{1}{c|}{0.64$\pm$0.08} & \multicolumn{1}{c|}{0.64$\pm$0.04} & 0.64$\pm$0.04 \\ \hline
        0.85--0.90                         & \multicolumn{1}{c|}{0.66$\pm$0.07} & \multicolumn{1}{c|}{0.63$\pm$0.06} & 0.65$\pm$0.06 \\ \hline
        0.90--0.95                         & \multicolumn{1}{c|}{0.68$\pm$0.10}  & \multicolumn{1}{c|}{0.67$\pm$0.10}  & 0.68$\pm$0.10  \\ \hline
        \end{tabular}
        \caption{Fraction of femtoscopically correlated pairs in 0-5\% to 20-30\% centrality bins for different $\beta_{\rm T}$ bins. \label{tab:fG_bT_1}}        
        \end{table}
        
        \begin{table}[!h]
        \begin{tabular}{|c|cc|}
        \hline
        \multirow{2}{*}{$\beta_{\rm T}$} & \multicolumn{2}{c|}{\begin{tabular}[c]{@{}c@{}}Pair-purity for \\ different centralities\end{tabular}} \\ \cline{2-3} 
                                         & \multicolumn{1}{c|}{30-40\%}                                 & 40-50\%                                 \\ \hline
        0.70--0.75                         & \multicolumn{1}{c|}{0.69$\pm$0.07}                           & 0.64$\pm$0.02                           \\ \hline
        0.75--0.80                         & \multicolumn{1}{c|}{0.65$\pm$0.05}                           & 0.60$\pm$0.07                           \\ \hline
        0.80--0.85                         & \multicolumn{1}{c|}{0.60$\pm$0.09}                           & 0.62$\pm$0.05                           \\ \hline
        0.85--0.90                         & \multicolumn{1}{c|}{0.61$\pm$0.07}                           & 0.62$\pm$0.08                           \\ \hline
        0.90--0.95                         & \multicolumn{1}{c|}{0.67$\pm$0.10}                            & 0.68$\pm$0.10                          \\ \hline
        \end{tabular}
        \caption{Fraction of femtoscopically correlated pairs in 30-40\% to 40-50\% centrality bins for different $\beta_{\rm T}$ bins. \label{tab:fG_bT_2}}
        \end{table}

\begin{figure}[!h]
            \centering    {\includegraphics[width=0.5\textwidth]{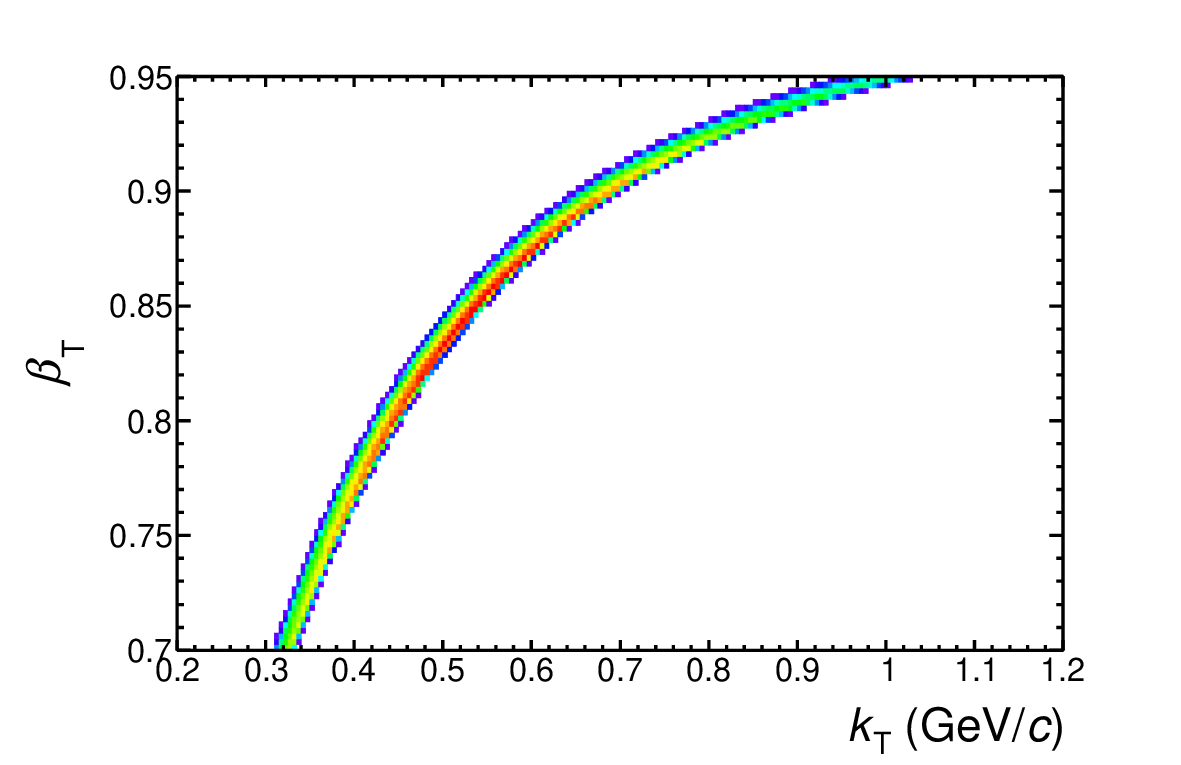}}
            \caption{Pair $\beta_{\mathrm{T}}$ distribution as a function of $k_{\mathrm{T}}$.}
            \label{fig:bT_kT}
        \end{figure}

It is also crucial to quantify $f_G$ in the different ranges of $k_{\mathrm{T}}$ for investigating the dependence of the source parameters on $\langle k_{\mathrm{T}}\rangle$. Instead of calculating from scratch, the $f_G$ value in each $k_{\mathrm{T}}$ bin is estimated in the following way. The $\beta_{\mathrm{T}}$ distribution is studied as a function of $k_{\mathrm{T}}$ as shown in Fig.~\ref{fig:bT_kT}. The weighted average of $\beta_{\mathrm{T}}$ is estimated from this distribution in each of the following $k_{\mathrm{T}}$ bins: 0.3$-$0.4, 0.4$-$0.5, 0.5$-$0.6, 0.6$-$0.8 and 0.8$-$1.0 GeV$/c$, which are listed in Table~\ref{tab:bTkT}. Then to estimate $f_G$ value in a certain $k_{\mathrm{T}}$ bin, one can refer to the corresponding $\beta_{\mathrm{T}}$ range from Tables~\ref{tab:fG_bT_1} and \ref{tab:fG_bT_2}. These values were used in the final fitting procedure to obtain the results shown in the next section. They can also be used as correction factors in true experimental analyses of pion-kaon correlations in Pb--Pb collisions at LHC energies.

        \begin{table}[!h]
        \begin{tabular}{|c|c|}
        \hline
        \multirow{2}{*}{$k_{\rm T}$ (GeV/$c$)} & \multirow{2}{*}{$\langle\beta_{\mathrm{T}}\rangle$} \\
                                               &                                                     \\ \hline
        0.3--0.4                                & 0.74                                            \\ \hline
        0.4--0.5                                & 0.81                                            \\ \hline
        0.5--0.6                                & 0.86                                            \\ \hline
        0.6--0.8                                & 0.90                                            \\ \hline
        0.8--1.0                                & 0.94                                            \\ \hline
        \end{tabular}
        \caption{$\langle\beta_{\rm T}\rangle$ in different $k_{\mathrm{T}}$ bins.} \label{tab:bTkT}        
        \end{table}

                
\section{Results} 
\label{sec:results}

This section presents our results on the source parameters extracted from the fits to femtoscopic correlation functions calculated in LQTH and iHKM models, as well as their dependence on the event multiplicity and the pair transverse velocity. 

         \begin{figure*}[htbp]
				 \centering
         {\includegraphics[width=0.8\textwidth]{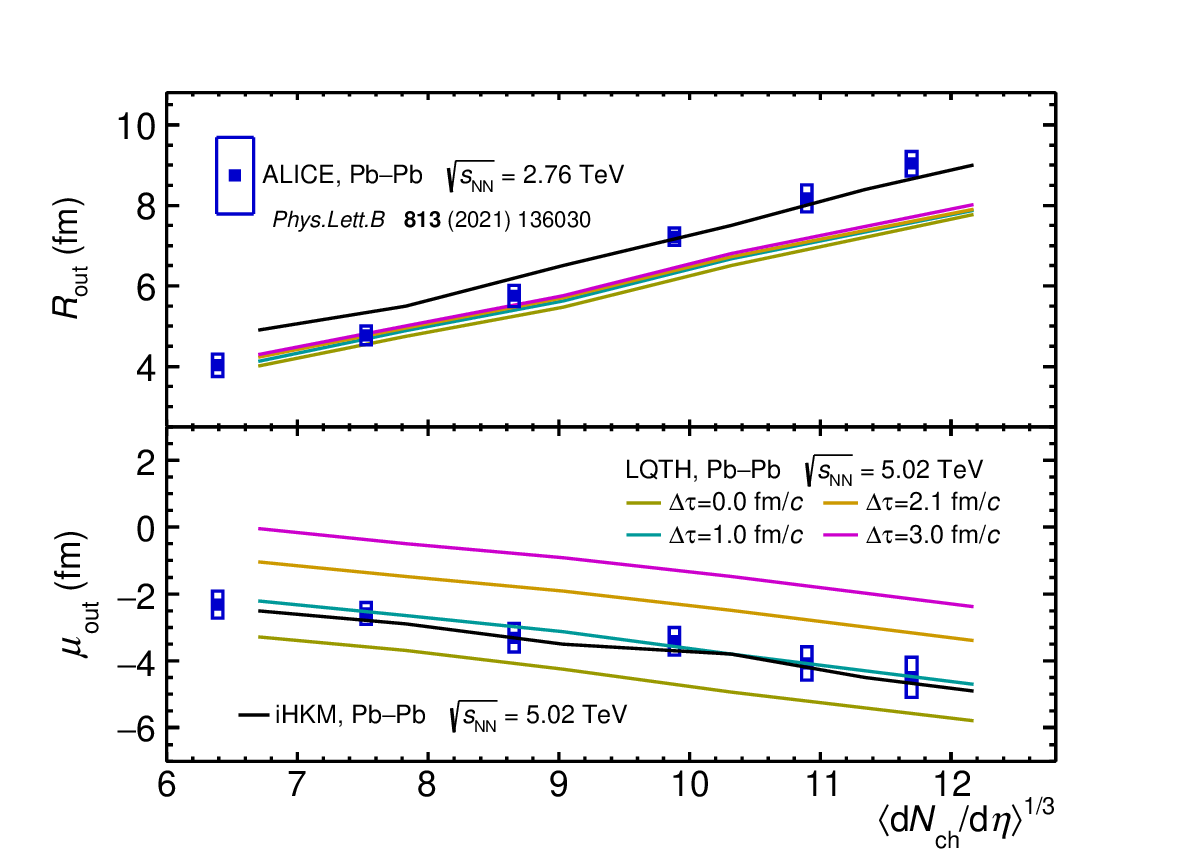}}
				 \caption{(Color online) $R_{\rm out}$ (upper panel) and $\mu_{\rm out}$ (lower panel) as functions of $\langle \mathrm{d}N_{\rm ch}/\mathrm{d}\eta\rangle^{\frac{1}{3}}$ for $\sqrt{s_{\rm NN}}=5.02$ TeV Pb$-$Pb collisions simulated using LQTH and iHKM. Different colors in LQTH results correspond to calculations with different values of additional delays in kaon emission.}
				  \label{fig:R_mu_int_therm}
		\end{figure*}

In Fig.~\ref{fig:R_mu_int_therm}, the $R_{\mathrm{out}}$ and $\mu_{\mathrm{out}}$ (solid lines) extracted from the correlation functions of charged pion-kaon pairs sampled in a wide 
pair $\beta_{\mathrm{T}}$ range are shown as functions of average charged particle multiplicity density $\langle \mathrm{d}N_{\rm ch}/\mathrm{d}\eta\rangle^{\frac{1}{3}}$. The values of primary fractions used for the fitting of these correlation functions are taken from~\cite{adam_nonid}. 
To effectively account for hadronic rescatterings at the post-hydrodynamic stage of collision in LQTH, where explicit microscopic treatment of this stage is not implemented, one can follow the technique used in~\cite{Therminator_Nonidentical_2760MeV}.
Namely, one can artificially introduce an additional time delay ($\Delta\tau$) between the pion and kaon emission in the model, which works as a proxy for rescattering. Here, along with the default configuration of no additional delay before kaon emission ($\Delta\tau=0$), three more cases with $\Delta\tau=1.0$, 2.1 and 3.0~fm$/c$ are considered for all centrality classes. The predictions are also compared with the results from pion-kaon femtoscopic analysis in Pb$-$Pb collisions at $\sqrt{s_{\rm NN}}=2.76$ TeV carried out by the ALICE Collaboration. The source sizes are assumed to follow a universal linear scaling on $\langle \mathrm{d}N_{\rm ch}/\mathrm{d}\eta\rangle^{\frac{1}{3}}$, regardless of the beam energy.


The calculated values of $R_{\mathrm{out}}$ in both LQTH and iHKM models are observed to increase linearly with increasing multiplicity. This type of behavior was already observed for identical pion and kaon femtoscopy scales~\cite{pion_2760,kaon_2760}, as well as in non-identical pion-kaon femtoscopy analysis~\cite{pionkaon_2760_alice} at LHC energies. However, a distinct offset between the values of the two models is clearly visible. In case of LQTH, the radii corresponding to different $\Delta\tau$'s for a given multiplicity do not differ from each other significantly, suggesting towards the trivial effect of the hadronic rescattering on the source radii. The LQTH predictions are consistent with the ALICE results in the lower multiplicity region but underestimate the data at higher multiplicities, whereas the iHKM predictions agree with the ALICE results at higher multiplicities and overestimate them at lower multiplicities. The values of $\mu_{\mathrm{out}}$ are observed to be universally negative, suggesting that the pions are always emitted closer to the center of the source and/or later than kaons. Moreover, the LQTH predictions with $\Delta\tau=1.0$ fm$/c$ match the iHKM predictions as well as the ALICE results providing a clear implication of the effect of hadronic rescattering on the emission asymmetry of the pairs along with the hydrodynamic flow. 
 
        \begin{figure*}[htbp]
				 \centering
         {\includegraphics[width=0.8\textwidth]{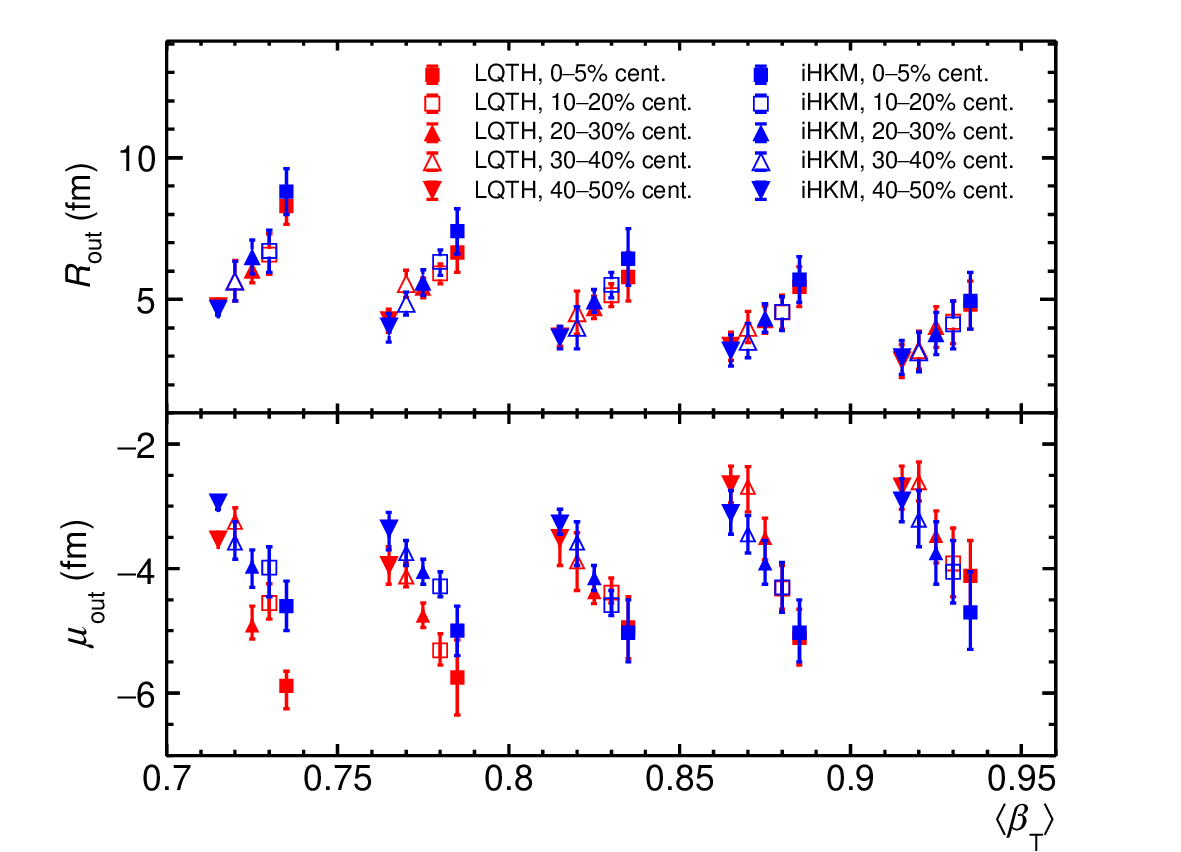}}
				 \caption{(Color online) $R_{\rm out}$ (upper panel) and $\mu_{\rm out}$ (lower panel) as functions of $\langle\beta_{\rm T}\rangle$ in different centrality classes for $\sqrt{s_{\rm NN}}=5.02$ TeV Pb$-$Pb collisions simulated using LQTH and iHKM. The points are shifted around each $\langle\beta_{\rm T}\rangle$ bin center for better visibility.}
				  \label{fig:R_mu_therm_ihkm}
		\end{figure*}

The dependence of $R_{\mathrm{out}}$ and $\mu_{\mathrm{out}}$ on $\langle\beta_{\mathrm{T}}\rangle$ for pion-kaon pairs in different centrality classes, obtained from the LQTH and iHKM models, are shown in Fig.~\ref{fig:R_mu_therm_ihkm}. 
For better visibility, the points are slightly shifted around each $\langle\beta_{\mathrm T}\rangle$ bin center. 
The LQTH source parameters discussed below are extracted from correlation functions simulated without incorporating an additional delay in kaon emission time. 
The uncertainties of the source parameters are estimated by fitting the correlation functions using the upper and lower limit of the primary pair fractions listed in Tables~\ref{tab:fG_bT_1} and \ref{tab:fG_bT_2}. 
The extracted radii vary in the range from 8.8~fm (for $\beta_{\rm T}: 0.70-0.75$ bin in 0$-$5\% centrality iHKM events) to 2.25~fm (for $\beta_{\rm T}: 0.90-0.95$ bin in 40$-$50\% centrality LQTH events). 
Results from both models agree within the estimated uncertainties. 
The observed decrease of $R_{\mathrm{out}}$ with increasing $\langle\beta_{\mathrm{T}}\rangle$ is consistent with the previously reported reduction of homogeneity lengths with increasing pair transverse momentum. 
For $\mu_{\mathrm{out}}$, the values are universally negative, ranging from $-6.25$~fm (for $\beta_{\rm T}: 0.70-0.75$ bin in 0$-$5\% central LQTH events) to $-2.95$~fm (for $\beta_{\rm T}: 0.85-0.90$ bin in 40$-$50\% central LQTH events). 
It exhibits a non-monotonic dependence and a decreasing trend on $\langle\beta_{\mathrm{T}}\rangle$, which is consistent with the decrease observed in the radii, with a maximum around $\beta_{\rm T}$ = 0.75--0.8.  
At higher $\langle\beta_{\mathrm{T}}\rangle$, the $\mu_{\mathrm{out}}$ values predicted by both models agree within uncertainties, whereas at lower $\langle\beta_{\mathrm{T}}\rangle$, the LQTH results show more negative $\mu_{\mathrm{out}}$ values than those from iHKM.
This discrepancy may stem from the absence of a hadronic rescattering phase in LQTH, leading to larger magnitudes of $\mu_{\mathrm{out}}$. 
The $\langle\beta_{\mathrm{T}}\rangle$ dependence of $\mu_{\mathrm{out}}$ predicted by LQTH is also slightly steeper than that obtained from iHKM. 
         \begin{figure*}[htbp]
				 \centering
             {\includegraphics[width=\textwidth]{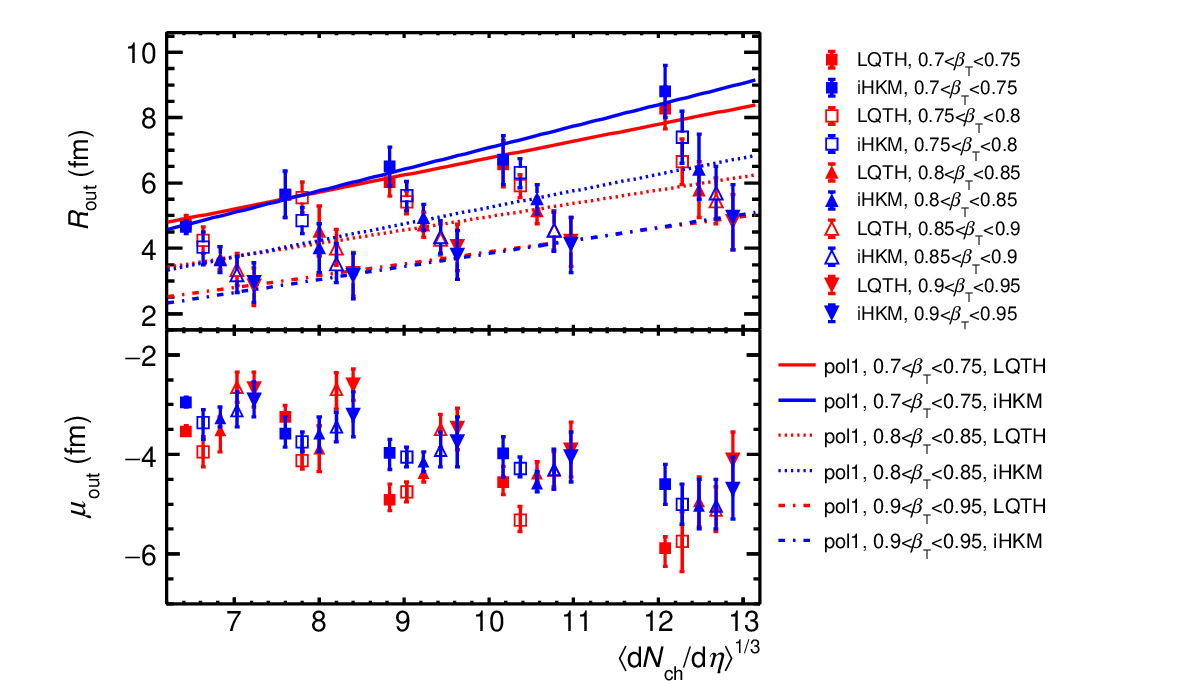}}
				 \caption{(Color online) $R_{\rm out}$ (upper panel) and $\mu_{\rm out}$ (lower panel) as functions of charged-particle multiplicity-density in different $\langle\beta_{\rm T}\rangle$ bins for $\sqrt{s_{\rm NN}}=5.02$ TeV Pb$-$Pb collisions simulated using LQTH and iHKM. The points are shifted around each $\langle \mathrm{d}N_{\rm ch}/\mathrm{d}\eta\rangle^{\frac{1}{3}}$ bin center for better visibility}
				  \label{fig:R_mu_therm_ihkm_cent}
		\end{figure*}

        \begin{figure*}[htbp]
				 \centering
         {\includegraphics[width=0.8\textwidth]{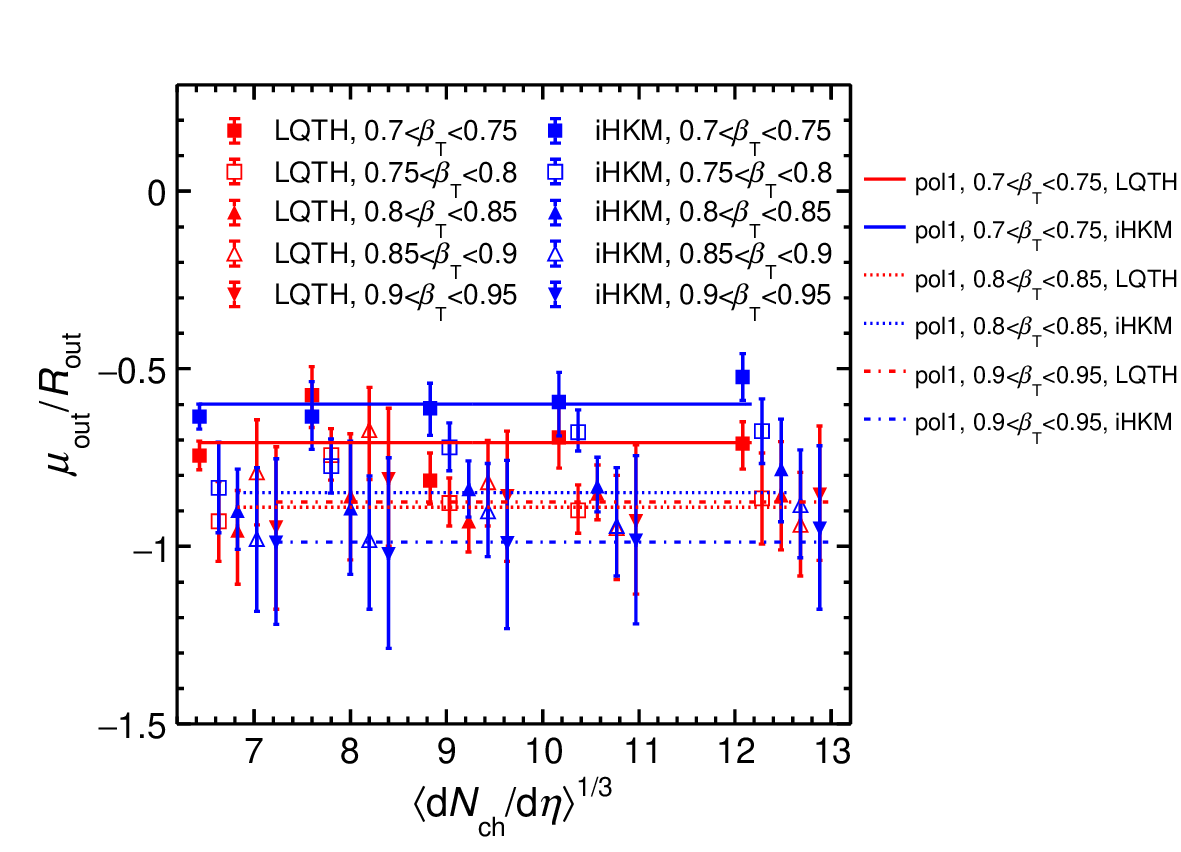}}
				 \caption{(Color online) The ratio of $\mu_{\rm out}$ to $R_{\rm out}$ as a function of charged-particle multiplicity density in different $\langle\beta_{\rm T}\rangle$ bins for $\sqrt{s_{\rm NN}}=5.02$ TeV Pb$-$Pb collisions simulated using LQTH and iHKM. The points are shifted around each $\langle \mathrm{d}N_{\rm ch}/\mathrm{d}\eta\rangle^{\frac{1}{3}}$ bin center for better visibility}
				  \label{fig:scaled_mu_therm_ihkm_cent}
		\end{figure*}
        
To observe the multiplicity-dependence of the source parameters more clearly, the $R_{\mathrm{out}}$ and $\mu_{\mathrm{out}}$ are plotted as functions of $\langle \mathrm{d}N_{\rm ch}/\mathrm{d}\eta\rangle^{\frac{1}{3}}$ in different $\beta_{\rm T}$ bins as shown in Fig.~\ref{fig:R_mu_therm_ihkm_cent} with the points shifted around each $\langle \mathrm{d}N_{\rm ch}/\mathrm{d}\eta\rangle^{\frac{1}{3}}$ bin center for better visibility. An expected monotonous decrease in the values of radii with the decreasing multiplicity for both model calculations is observed. 
Upon fitting the sets of predicted $R_{\mathrm{out}}$ points from both models for $\beta_{\rm T}:0.70-0.75, 0.80-0.85$ and $0.90-0.95$ ranges with a linear function, it is clearly seen that the radii from iHKM change more steeply than LQTH, which is consistent with the predictions of variation in the source-size with event multiplicity from the identical-pion femtoscopy using same model calculations for Pb$-$Pb collisions at $\sqrt{s_{\mathrm{NN}}}=5.02$~TeV~\cite{wiola}. 
In case of $\mu_{\mathrm{out}}$, the non-monotonous dependence on $\beta_{\rm T}$ is clearly visible in each multiplicity bin. 
As both  $R_{\mathrm{out}}$ and $|\mu_{\mathrm{out}}|$ decrease with increasing $\beta_{\rm T}$, it is essential to investigate the ratio of $\mu_{\mathrm{out}}$ to $R_{\mathrm{out}}$ in different $\beta_{\rm T}$ and $\langle \mathrm{d}N_{\rm ch}/\mathrm{d}\eta\rangle^{\frac{1}{3}}$ bins in order to quantify the effect of the collectivity of the system on the pair-emission dynamics, which is given in Fig.~\ref{fig:scaled_mu_therm_ihkm_cent}.  
The magnitude of scaled pair-emission asymmetry is observed to increase in the negative direction as the $\beta_{\rm T}$ increased from 0.70$-$0.75 to 0.90$-$0.95, implying that the kaons are pushed much farther from the center of the source due to the radial flow than pions and/or emitted much earlier than pions. 
Interestingly, the values of scaled $\mu_{\mathrm{out}}/R_{\mathrm{out}}$ for a given $\beta_{\rm T}$ interval are found to be similar, within the uncertainties, across all multiplicity bins.
This suggests that the relative asymmetry in particle emission depends on the relative size and position of the homogeneity regions, determined by the interplay between collective and thermal velocities, but not on the overall system size, which varies with collision centrality. 

\begin{figure}[!h]
				 \centering
         {\includegraphics[width=0.5\textwidth]{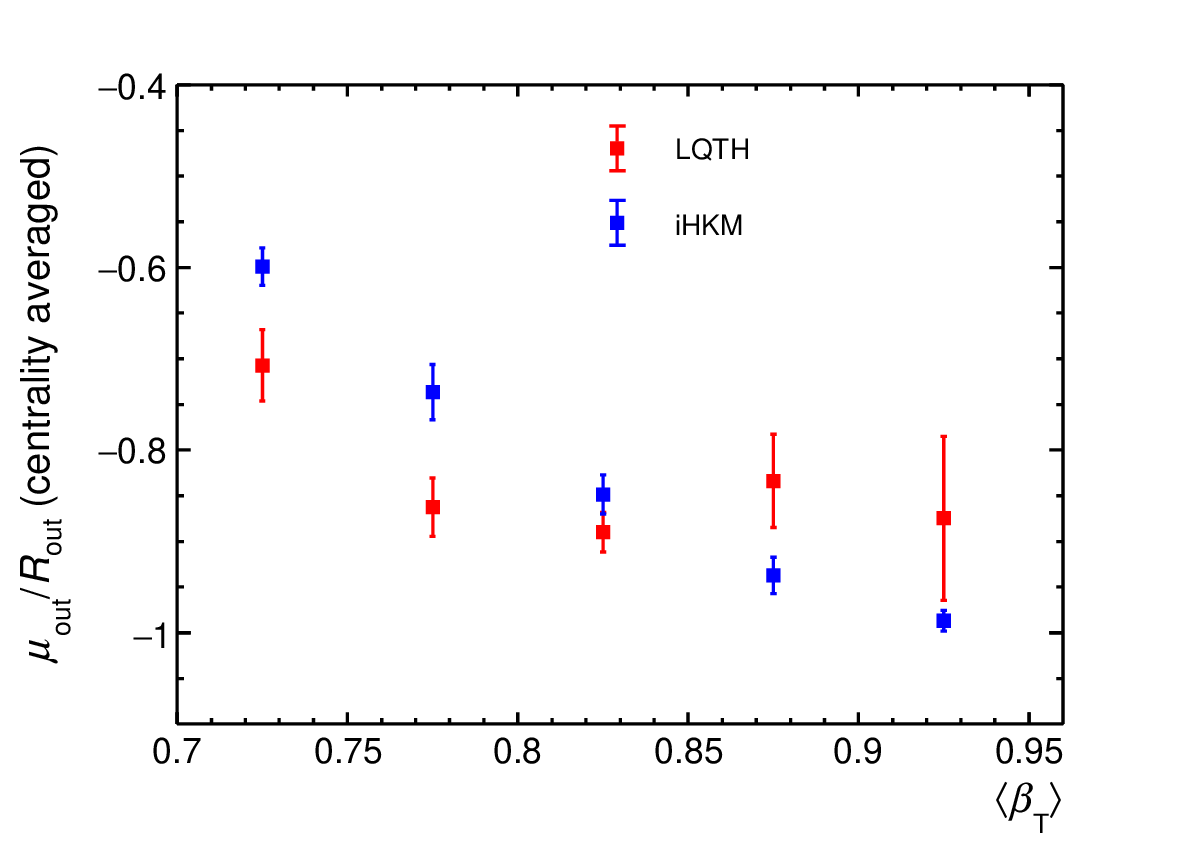}}
				 \caption{(Color online) The  centrality-averaged ratio of $\mu_{\rm out}$ to $R_{\rm out}$ as a function of $\langle\beta_{\rm T}\rangle$ bins for $\sqrt{s_{\rm NN}}=5.02$ TeV Pb$-$Pb collisions simulated using LQTH and iHKM.} \label{fig:avgd_scaled_mu_therm_ihkm_cent}
		\end{figure}

The scaled pair-emission asymmetry averaged over multiplicity bins are shown in Fig.~\ref{fig:avgd_scaled_mu_therm_ihkm_cent} as a function of $\langle\beta_{\mathrm{T}}\rangle$, where the iHKM results show a smooth increase in the magnitude over increasing $\langle\beta_{\mathrm{T}}\rangle$ values whereas the LQTH results show a non-monotonous trend.        

         \begin{figure}[!h]
				 \centering 
         {\includegraphics[width=0.5\textwidth]{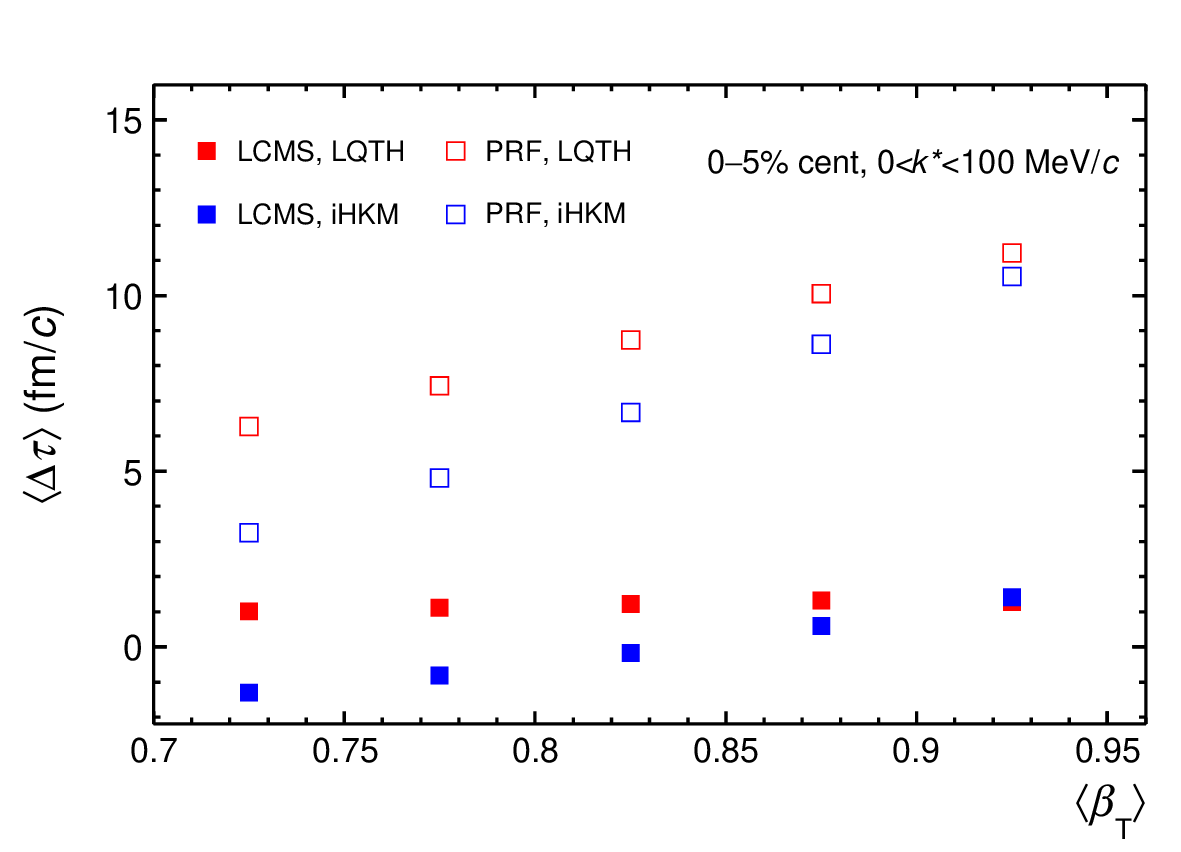}}
				 \caption{(Color online) The averaged difference between pion and kaon emission proper times in LCMS and PRF as a function of $\langle\beta_{\rm T}\rangle$ for $\sqrt{s_{\rm NN}}=5.02$ TeV Pb$-$Pb collisions simulated using LQTH and iHKM. 
                 } \label{fig:MeanDeltaT_LCMS}
		\end{figure}

The negative sign of $\mu_\mathrm{out}$ in iHKM
suggests that pions are emitted later (and/or closer to the system's center) than kaons. This result apparently contradicts the one
obtained earlier within HKM in~\cite{mtscales1} (see also~\cite{mtscales2,mtscales3}) and then confirmed in ALICE experimental analysis~\cite{kaon_2760} 
for maximal emission proper times of pions and kaons, 
$\tau_\pi$ and $\tau_K$, estimated based on simultaneous fitting of 
pion and kaon $p_T$ spectra and femtoscopy radii $R_\mathrm{long}$.
A similar relation between $\tau_\pi$ and $\tau_K$ is obtained
in iHKM if one estimates it as average values of respective
emission time distributions~\cite{mtscales3}.
The reason for the discrepancy between mentioned result and the result of the present analysis should be studied separately,
since these two results were obtained within two different approaches
to the pion-kaon emission asymmetry estimation. Namely, in previous
iHKM studies one determined the difference between effective proper times $\tau$ characterizing the emission of the two considered particle species in LCMS. In the current analysis we fit the two-particle correlation functions in PRF and characterize both spatial
and temporal asymmetry between pion and kaon emission by single parameter $\mu_\mathrm{out}$, entering the source function formula in the Gaussian approximation~\eqref{eq:sourcefcn}. The information about
times of particle emission is not explicitly reflected in the source function, since the latter is obtained from the emission function by integrating it over time.

Preliminary analysis of the pair emission time difference distributions in iHKM (see Fig.~\ref{fig:MeanDeltaT_LCMS}) suggests that the discrepancy between the time asymmetries obtained in the two approaches 
can be connected with the Lorentz transformation effect --- in PRF the average pion-kaon emission time differences in iHKM are positive at all 
$\beta_\mathrm{T}$,
in agreement with the obtained results for $\mu_\mathrm{out}$. 
However, the same differences in LCMS are negative at low and intermediate 
$\beta_\mathrm{T}$, which agrees with the maximal emission times estimation in~\cite{mtscales1,mtscales2,mtscales3}. 

In the LQTH model the mean emission time differences are positive both in LCMS
and PRF reference frames. The observed difference in LCMS $\langle\Delta \tau\rangle$ sign between the iHKM and the LQTH models can be connected
with the simulation of the hadronic stage of collision, 
present in iHKM and absent in LQTH.
In particular, the previously mentioned recombination and subsequent decay of $K^*$ resonances implemented in UrQMD model, which describes the afterburner stage in iHKM, can lead to later kaon emission times and thus negative $\langle\Delta \tau\rangle$ values.
We think that clarifying the role of various factors in formation of
the observed emission picture requires more thorough investigation in future studies.

\section{Conclusions} 
\label{sec:conclusion}

In this work, two models with substantially different treatments of the hadronic phase were employed to predict the differences in the first moment of the emission distribution of pions and kaons (the emission asymmetry--$\mu_{\rm{out}}$) in the out direction, along the pair transverse momentum. 
This comparison enables the assessment of the effects of pion--kaon interactions, resonance regeneration and rescattering, and the evaluation of the sensitivity of the corresponding femtoscopic observable.

The analysis demonstrates that the emission asymmetry is sensitive to time and spatial shifts of the order of 1~fm/$c$, which corresponds to characteristic timescales of the hadronic evolution. 
Such sensitivity indicates that even relatively short-lived processes occurring during the hadronic phase—such as delayed kaon emission due to rescattering or resonance regeneration—can produce measurable effects in the femtoscopic observables. 
Therefore, this observable serves as a precise probe of the space–time structure of the emission function and the duration of the hadronic stage.

A non-monotonic behavior of the parameter $\mu_{\rm{out}}$ is observed, reflecting the complex interplay between regeneration and decay of the K* resonance as a function of the pair transverse velocity. 
This feature underlines the capability of the observable to constrain the relative contribution of competing hadronic processes and makes it particularly suitable for fine-tuning of model parameters when compared with experimental data.
Furthermore, the scaled emission asymmetry $\mu_{\rm{out}}$/$R_{\rm{out}}$ for pion-kaon pairs appears to be independent of particle multiplicity. It remains to be investigated whether this behavior results from a specific interplay of processes unique to these two models and this collision system, or whether it represents a more general scaling relation. 

In addition to this, it was found that the average proper time distance between pions and kaons $\langle \Delta \tau\rangle$ in LCMS frame, when analyzed as a function of the transverse velocity, shows a flip of the sign when the full physical description of the hadronic phase is switched on. This effect is not observed when a sudden freeze-out scenario is implemented, nor when the system is analysed using the PRF frame. 

Finally, the study confirms that the appropriate scaling variable for both $R_{\rm{out}}$ and $\mu_{\rm{out}}$ is the particle multiplicity, emphasizing the universal character of the observed dependencies across different collision energies.

In summary, this work demonstrates the relevance of the hadronic phase in the evolution of relativistic heavy ion collisions and their impact on the emission asymmetries between particles with different masses, such as pions and kaons. The work provides predictions for $\sqrt{s_{\mathrm{NN}}}=5.02$~TeV Pb$-$Pb collisions.  In addition to the interactions, it is worth noting that the chosen reference frame -- PRF of LCMS -- also influences the conclusions of which of the particle species decouples from the system earlier or later. This aspect has to be explored further in future works.

\begin{acknowledgments}
Studies of Yu.S. were funded by IDUB-POB projects granted by WUT (Excellence Initiative: Research University (ID-UB)). The work was also supported by a grant from the Simons Foundation (Grant SFI-PD-Ukraine-00014578, Award number 1290596, V.S.). This work was supported by the Polish National Science Centre under decision no. UMO-2022/45/B/ST2/02029. This work was also supported by the Research University – Excellence Initiative of Warsaw University of Technology via the strategic funds of the Priority Research Centre of High Energy Physics and Experimental Techniques and the IDUB POSTDOC programme. This work was also supported by the Department of Science and Technology (DST), Government of India.
\end{acknowledgments}


\begin{thebibliography}{50}
\bibitem{qgp1}
J. Adams \textit{et al.} (STAR Collaboration), 
\textit{Nucl. Phys. A} \textbf{757} 102--183 (2005).

\bibitem{qgp2}
K. Adcox \textit{et al.} (PHENIX Collaboration), 
\textit{Nucl. Phys. A} \textbf{757} 184--283 (2005).

\bibitem{goldhaber}
G. Goldhaber, S. Goldhaber, W. Lee, and A. Pais, 
\textit{Phys. Rev.} \textbf{120}, 300 (1960).

\bibitem{kopylov1}
G. I. Kopylov and M. I. Podgoretsky, 
\textit{Sov. J. Nucl. Phys.} \textbf{15}, 219 (1972).

\bibitem{kopylov2}
G. I. Kopylov, 
\textit{Phys. Lett. B} \textbf{50}, 472--474 (1974).

\bibitem{cocconi}
G. Cocconi, 
\textit{Phys. Lett. B} \textbf{49}, 459--461 (1974).

\bibitem{gyulassy}
M. Gyulassy, S. K. Kauffmann, and L. W. Wilson, 
\textit{Phys. Rev. C} \textbf{20}, 2267 (1979).

\bibitem{pratt-hbt} 
S. Pratt, 
\textit{Phys. Rev. D} {\bf  33}, 1314 (1986).

\bibitem{makhsin87}
A. N. Makhlin and Yu. M. Sinyukov, 
\textit{Yad. Fiz.} \textbf{46}, 637 (1987).

\bibitem{makhsin88}
A. N. Makhlin and Yu. M. Sinyukov, 
\textit{Z. Phys. C} \textbf{39}, 69 (1988).

\bibitem{sin89} 
Yu. M. Sinyukov, 
\textit{Nucl. Phys. A} {\bf 498}, 151--159 (1989).

\bibitem{hama-hbt} 
Y. Hama, S. S. Padula, 
\textit{Phys. Rev. D} {\bf 37}, 3237 (1988).

\bibitem{sin94}
Yu. M. Sinyukov, 
\textit{Nucl. Phys. A} \textbf{566}, 589 (1994).

\bibitem{akksin95}
S. V. Akkelin, Yu. M. Sinyukov, 
\textit{Phys. Lett. B} \textbf{356}, 525 (1995).

\bibitem{kaon_2760}
S.~Acharya \textit{et al.} (ALICE Collaboration), \textit{Phys.~Rev.~C} {\bf 96}, 064613 (2017).



\bibitem{hkm-kaon}
V. M. Shapoval, P. Braun-Munzinger, Iu. A. Karpenko, Yu. M. Sinyukov,
\textit{Nucl. Phys. A} \textbf{929}, 1--8 (2014).


\bibitem{led1}
R. Lednicky, V. L. Lyuboshits, B. Erazmus, and D. Nouais, 
\textit{Phys. Lett. B} \textbf{373} 30--34 (1996).

\bibitem{led2}
S. Voloshin, R. Lednicky, S. Panitkin, and N. Xu, 
\textit{Phys. Rev. Lett.} \textbf{79} 4766--4769, (1997).

\bibitem{adam_nonid}
A.~Kisiel, \textit{Phys. Rev. C} {\bf 81}, 064906 (2010).

\bibitem{mtscales1}
Yu. M. Sinyukov, V. M. Shapoval, V. Yu. Naboka, 
\textit{Nucl. Phys. A} \textbf{946}, 227--239 (2016).

\bibitem{mtscales2}
V.~M.~Shapoval, Yu.~M.~Sinyukov, 
\textit{Nucl. Phys. A} \textbf{1016}, 122322 (2021).

\bibitem{mtscales3}
Y.~Sinyukov, V.~Shapoval, M.~Adzhymambetov, 
\textit{Universe} \textbf{9}, 433 (2023).

\bibitem{kstar} V.~M.~Shapoval, P.~Braun-Munzinger, Yu.~M.~Sinyukov, 
\textit{Nucl. Phys. A} \textbf{968}, 391--402 (2017).

\bibitem{pionkaon_2760_alice}
S.~Acharya \textit{et al.} (ALICE Collaboration), \textit{Phys.~Rev.~Lett.} {\bf{813}}, 136030 (2021).

\bibitem{ihkm2}
V.~Yu.~Naboka, S.~V.~Akkelin, Iu.~A.~Karpenko, Yu.~M.~Sinyukov,  
\textit{Phys. Rev. C} \textbf{91}, 014906 (2015).

\bibitem{ihkm} 
V.~Yu.~Naboka, Iu.~A.~Karpenko, Yu.~M.~Sinyukov, 
\textit{Phys. Rev. C} \textbf{93}, 024902 (2016).

\bibitem{gliss}
P.~Bozek, W.~Broniowski, M.~Rybczynski, G.~Stefanek, 
\textit{Comput. Phys. Commun.} \textbf{245}, 106850 (2019).

\bibitem{cgc1}
F. Gelis, E. Iancu, J. Jalilian-Marian, R. Venugopalan, 
\textit{Ann. Rev. Nucl. Part. Sci.} \textbf{60}, 463 (2010).

\bibitem{cgc2}
E. Iancu, A. Leonidov, L. McLerran, 
arXiv:hep-ph/0202270.

\bibitem{is3}
W.~Israel, J.~M.~Stewart, 
\textit{Ann. Phys.} \textbf{118} 341--372 (1979).

\bibitem{hlle}
Iu. Karpenko, P. Huovinen, M. Bleicher, 
\textit{Comput. Phys. Commun.} \textbf{185}, 3016 (2014).

\bibitem{laine}
M. Laine, Y. Schroeder, 
\textit{Phys. Rev. D} \textbf{73}, 085009 (2006).

\bibitem{hotqcd}
A. Bazavov \textit{et al.} (The HotQCD Collaboration), 
\textit{Phys. Rev. D} \textbf{90}, 094503 (2014).

\bibitem{cooper-frye}
F. Cooper and G. Frye, 
\textit{Phys. Rev. D} \textbf{10}, 186 (1974).

\bibitem{grad} 
D. Molnar, 
\textit{J. Phys. G} \textbf{38}, 124173 (2011).

\bibitem{cornelius3} 
D. Molnar, Z. Wolff, 
\textit{Phys. Rev. C} \textbf{95}, 024903 (2017).

\bibitem{cornelius1} 
P. Huovinen, H. Petersen, 
\textit{Eur. Phys. J. A} \textbf{48}, 171 (2012).

\bibitem{urqmd1}
S.~A.~Bass \textit{et al.}, 
\textit{Prog. Part. Nucl. Phys.} \textbf{41} (1998) 225--370. 

\bibitem{urqmd2}
M.~Bleicher \textit{et al.}, 
\textit{J. Phys. G} \textbf{25} (1999) 1859--1896.

\bibitem{lqth1}
P. Bozek and I. Wyskiel-Piekarska, 
\textit{Phys. Rev. C} \textbf{85}, 064915 (2012).

\bibitem{lqth2}
M. Chojnacki, A. Kisiel, W. Florkowski, and W. Broniowski,
\textit{Comput. Phys. Commun.} \textbf{183}, 746 (2012).

\bibitem{Therminator_identical_2760MeV}
A.~Kisiel, M.~Ga\l{}a\ifmmode \dot{z}\else \.{z}\fi{}yn, and P.~Bo\ifmmode \dot{z}\else \.{z}\fi{}ek, \textit{Phys.~Rev.~C} {\bf{90}}, 064914 (2014). 

\bibitem{source}
V.~M.~Shapoval, Yu.~M.~Sinyukov. Iu.~A.~Karpenko, 
\textit{Phys. Rev. C} \textbf{88}, 064904 (2013).

\bibitem{koonin0}
E.~Fermi, \textit{Z. Phys.} \textbf{88}, 161 (1934), translated in F.~L.~Wilson, \textit{Am. J. Phys.} \textbf{36}, 1150 (1968).

\bibitem{koonin1}
S.~E.~Koonin, \textit{Phys. Lett. B} \textbf{70}, 43 (1977).

\bibitem{koonin2}
R.~Lednicky and V.~L.~Lyuboshitz, \textit{Sov. J. Nucl. Phys.} \textbf{35}, 770 (1982).

\bibitem{koonin3}
S.~Pratt, T.~Csorgo, and J.~Zimanyi, \textit{Phys. Rev. C} \textbf{42},  2646 (1990).

\bibitem{Therminator_identical_5020MeV}
P.~Chakraborty, A.~K.~Pandey, and S.~Dash, \textit{Eur.~Phys.~J.~A} {\bf{57}}, 338 (2021).

\bibitem{Lednicky:2005tb}
R.~Lednicky,
Phys. Part. Nucl. \textbf{40}, 307--352 (2009).

\bibitem{adamsh}
A.~Kisiel, and D.~A.~Brown, \textit{Phys.~Rev.~C} {\bf{80}}, 064911 (2009).

\bibitem{Therminator_Nonidentical_2760MeV}
A.~Kisiel, \textit{Phys.~Rev.~C} {\bf 98}, 044909 (2018).

\bibitem{pion_2760}
J.~Adam \textit{et al.} (ALICE Collaboration), \textit{Phys.~Rev.~C} {\bf 93}, 024905 (2016).

\bibitem{wiola}
W.~Rzesa, G.~Kornakov, A.~R.~Kisiel, Yu.~M.~Sinyukov, and V.~M.~Shapoval, \textit{Phys.~Rev.~C} {\bf{110}}, 034904 (2024).









 
\end{thebibliography}
\end{document}